\documentclass[10pt,twocolumn,letterpaper]{article}

\usepackage[pagenumbers]{iccv} 
\usepackage{multirow}

\definecolor{iccvblue}{rgb}{0.21,0.49,0.74}
\usepackage[pagebackref,breaklinks,colorlinks,allcolors=iccvblue]{hyperref}

\def\paperID{*****} 
\def\confName{ICCV}
\def\confYear{2025}

\title{TAGF: Time-aware Gated Fusion for Multimodal Valence-Arousal Estimation}

\author{Yubeen Lee\\
Sungkyunkwan University\\
\texttt{lybin1070@gmail.com}\\
\and
Sangeun Lee\\
Electronics and Telecommunications Research Institute \\
\texttt{sange1104@etri.re.kr}\\
\and
Chaewon Park\\
Sungkyunkwan University \\
\texttt{chaewon24caley@gmail.com}\\
\and
Junyeop Cha\\
Sungkyunkwan University \\
\texttt{jycha95@g.skku.edu}\\
\and
Eunil Park\thanks{Corresponding author}\\
Sungkyunkwan University \\
\texttt{eunilpark@skku.edu}\\}

\begin{document}
\maketitle
\begin{abstract}
Multimodal emotion recognition often suffers from performance degradation in valence-arousal estimation due to noise and misalignment between audio and visual modalities. To address this challenge, we introduce TAGF, a Time-aware Gated Fusion framework for multimodal emotion recognition. The TAGF adaptively modulates the contribution of recursive attention outputs based on temporal dynamics. Specifically, the TAGF incorporates a BiLSTM-based temporal gating mechanism to learn the relative importance of each recursive step and effectively integrates multistep cross-modal features. By embedding temporal awareness into the recursive fusion process, the TAGF effectively captures the sequential evolution of emotional expressions and the complex interplay between modalities. Experimental results on the Aff-Wild2 dataset demonstrate that TAGF achieves competitive performance compared with existing recursive attention-based models. Furthermore, TAGF exhibits strong robustness to cross-modal misalignment and reliably models dynamic emotional transitions in real-world conditions.

\end{abstract}    
\begin{figure}[t]
  \centering
   \includegraphics[width=1.0\linewidth]{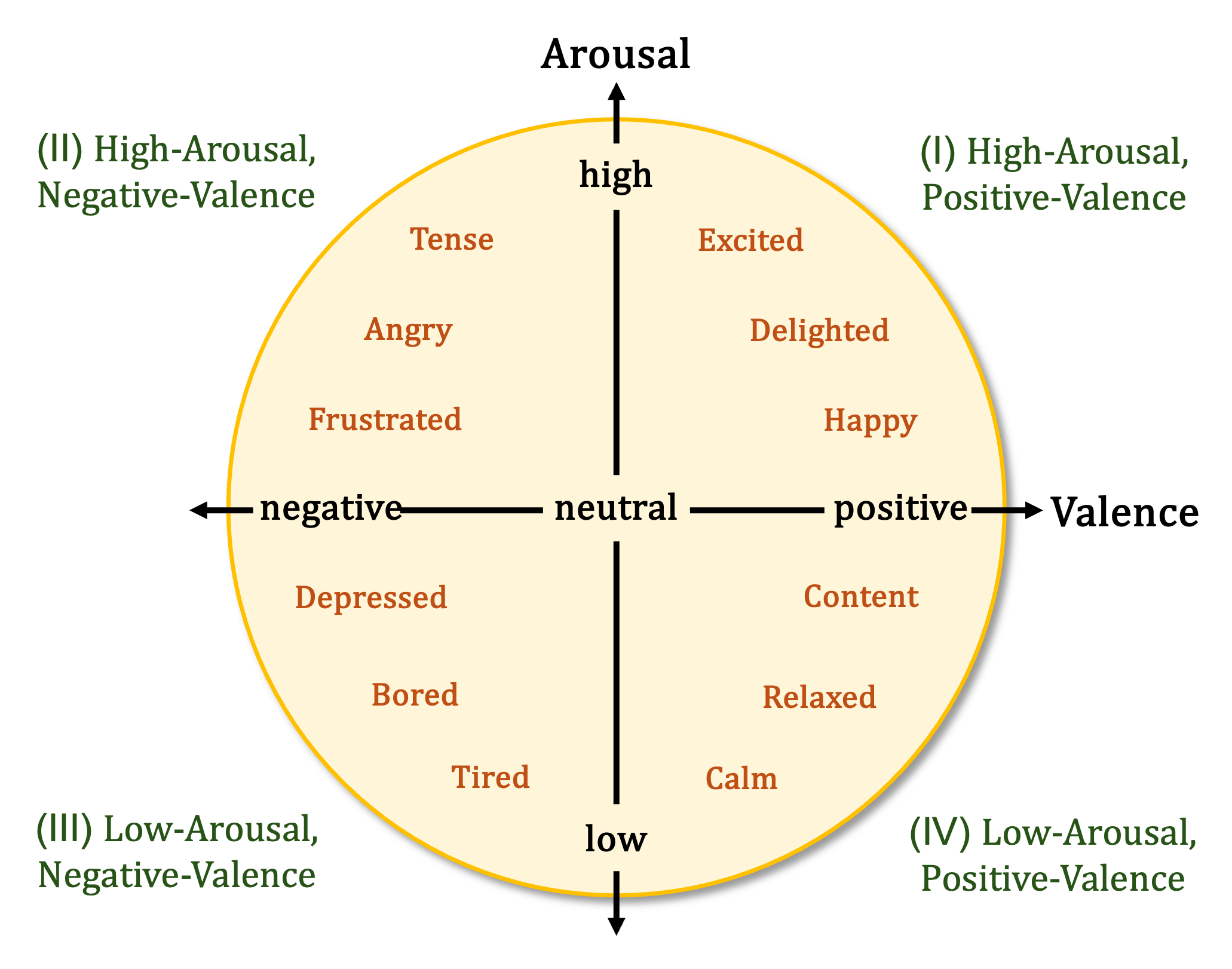}
   \caption{A two-dimensional Valence-Arousal space illustrating emotional states across four affective quadrants.}
   \label{fig:figure1}
\end{figure}

\section{Introduction}
\label{sec:intro}
Emotion recognition has become increasingly important in the field of machine intelligence, serving as the foundation for building empathetic AI systems~\cite{lagrandeur2015emotion}. While traditional methods often rely on categorical emotion classification~\cite{tripathi2017using,xie2019speech}, Valence-Arousal (VA) estimation represents emotions as continuous states in a two-dimensional space~\cite{feldman1995valence}. Valence refers to the degree of pleasantness, and arousal indicates its intensity, as depicted in Figure~\ref{fig:figure1}. This continuous representation enables the model to capture the subtle and ambiguous nature of human emotions, which are often difficult to handle effectively using categorical methods.

The 9th Affective Behavior Analysis in-the-Wild (ABAW) Competition builds upon a series of previous challenges~\cite{zafeiriou2017aff,kollias2020analysing,kollias2021analysing,kollias2022abaw,kollias2023abaw,kollias2023abaw2,kollias20246th,kollias20247th,kolliasadvancements}. Among its tasks, the competition includes VA estimation, which utilizes an extended version of the Aff-Wild2 database~\cite{kollias2019deep,kollias2019expression,kollias2021affect,kollias2023multi,kollias2025dvd}, an audio-visual dataset collected in unconstrained, real-world settings. The VA estimation task aims to predict the frame-level valence and arousal scores of individuals appearing in video clips by leveraging multimodal audio-visual cues. 



Recognizing emotions in naturalistic settings requires models that can handle the dynamic nature of emotional expression. Because emotions evolve continuously over time, modeling temporal variability is crucial for accurate predictions. However, previous studies based on Joint Cross-Attention (JCA) architectures~\cite{praveen2023recursive,praveen2024recursive} primarily focused on integrating information across modalities using fixed cross-attention weights without considering the temporal context. These approaches apply the same gating pattern uniformly across all time steps, limiting their ability to adapt to the temporal dynamics of the emotional states.

To address this issue, we propose a Time-aware Gated Fusion (TAGF) model that dynamically adjusts the importance of each attention layer over time. Specifically, we regard the outputs of the cross-attention layers as a temporal sequence and apply a temporal encoder to dynamically adjust the importance of the frame-level cross-attention representations according to their temporal relevance. This allows the model to better capture the dynamics of emotion by selectively focusing on informative cues while ignoring irrelevant or noisy signals. Experimental results demonstrate that the proposed model achieves competitive performance compared to existing methods on in-the-wild VA estimation benchmarks.

\section{Related Work}
\label{sec:related work}

\subsection{Valence-Arousal Estimation}
VA estimation is based on Russell's Circumplex Model~\cite{russell1980circumplex}, which represents emotions in a continuous two-dimensional space. This approach enables a more nuanced and precise understanding of emotional states than traditional categorical emotion classification methods.

Early research primarily relied on handcrafted features extracted from facial images or audio signals, which were then fed into conventional machine learning models~\cite{meng2022valence, zhang2021continuous}. However, these approaches are often constrained by the inherent complexity and subjectivity of manual feature extraction.

The development of deep learning has significantly advanced this field, with Convolutional Neural Networks (CNNs) becoming a foundational component of VA estimation systems~\cite{hwooi2022deep}. \citet{zhang2020m} proposed a CNN-RNN framework that extracts spatiotemporal features from video using either 3D CNNs or pre-trained 2D CNNs, which are then integrated with Bidirectional RNNs. Several approaches have combined CNNs with LSTMs to better capture the temporal dynamics of facial expressions by independently processing visual and audio modalities and subsequently integrating them using a late-fusion strategy~\cite{karas2022continuous}.

Recently, Transformer-based multimodal approaches have gained increasing attention for VA estimation. MAVEN~\cite{ahire2025maven} introduced a framework in which transformers utilize attention mechanisms to effectively integrate information from multiple modalities, including facial expressions, audio, and text. \citet{ming2022cnn} proposed a CNN-LSTM-based method for facial expression recognition that employs a dual-layer attention mechanism to focus on the most emotionally important regions of the face.

\subsection{Multimodal Fusion and Attention Mechanism}
Multimodal emotion recognition has attracted increasing attention owing to its ability to enhance recognition accuracy by leveraging complementary information from the visual and auditory modalities. Recent studies explored various attention mechanisms and fusion strategies to enable the more effective integration of cross-modal features.

\citet{sanchez2013audiovisual} developed a three-stage fusion framework that combines regression outputs from audio and visual features to continuously estimate valence and arousal dimensions, following Russell’s circumplex model of affect. \citet{liu2023multi} introduced a multimodal fusion technique utilizing a saliency-based attention network to assign modality-specific weights, along with a complementary attention network to further refine feature integration. \citet{sharafi2022novel} proposed a hybrid multimodal fusion framework that extracts spatio-temporal features from video and MFCC-based features from audio, and integrates them using a BiLSTM network to enhance emotion classification performance. \citet{xie2023multimodal} presented a multimodal emotion recognition approach based on multitask learning and attention-enhanced fusion.

\citet{mocanu2023multimodal} introduced a method that incorporates multi-level attention into a visual 3D CNN, along with audio-visual cross-attention fusion to effectively model inter-modal relationships. \citet{zhang2024deep} proposed a temporal attention-based fusion framework that dynamically assigns importance weights to audio and video modalities over time for improved emotion recognition. \citet{ghaleb2023joint} designed an attention-based emotion recognition framework that leverages modality-specific Transformer encoders to focus on temporally important segments of audio and video. Additionally, \citet{praveen2022joint, praveen2023audio, praveen2024recursive} proposed a cross-attention mechanism to learn the complementary relationships between audio and visual inputs, extending this approach through joint feature representation and a recursive attention framework.

\subsection{Gated Attention for Multimodal Fusion}
Gating mechanisms play an essential role in multimodal learning by selectively controlling the flow of information~\cite{arevalo2020gated}. \citet{zhang2025multimodal} proposed the MSFN model to address the issue of feature misalignment that often arises in attention-based image-text sentiment analysis. Their approach involves learning the interactions between aligned features using a graph-based structure and applying a gated attention mechanism to suppress irrelevant alignments while selectively integrating meaningful sentiment-related information.

\citet{praveen2024incongruity} highlighted the challenges of modality degradation or weak cross-modal complementarity in audio-visual emotion recognition. To address these issues, they introduced the Incongruity-Aware Cross-Attention (IACA) model, which incorporates a dual-gating mechanism to effectively handle incongruent relationships. \citet{kumar2020gated} proposed a selective fusion strategy based on gating functions to mitigate the influence of noisy modalities in cross-modal interactions. Their method improves the robustness of multimodal emotion recognition by focusing on cross-modal information when beneficial and relying on unimodal cues when necessary.

Additionally, \citet{praveen2023recursive,praveen2024recursive} proposed the Gated Recursive Joint Cross-Attention (GRJCA) model, which integrates a gating mechanism into the Recurrent Joint Cross-Attention (RJCA) framework. This model dynamically adjusts the importance of each recurrent step, enabling the system to emphasize informative features, even under weak complementarity conditions. However, the original GRJCA model statically combined the outputs from all attention layers, limiting its ability to capture temporal dynamics or reflect evolving emotional states.

To overcome this limitation, the present study introduces Time-aware Gated Fusion, which dynamically reflects frame-level importance in response to temporal variations in emotional states.

\begin{figure*}[!ht]
  \centering
  \hfuzz=20pt
  \includegraphics[width=1\linewidth]{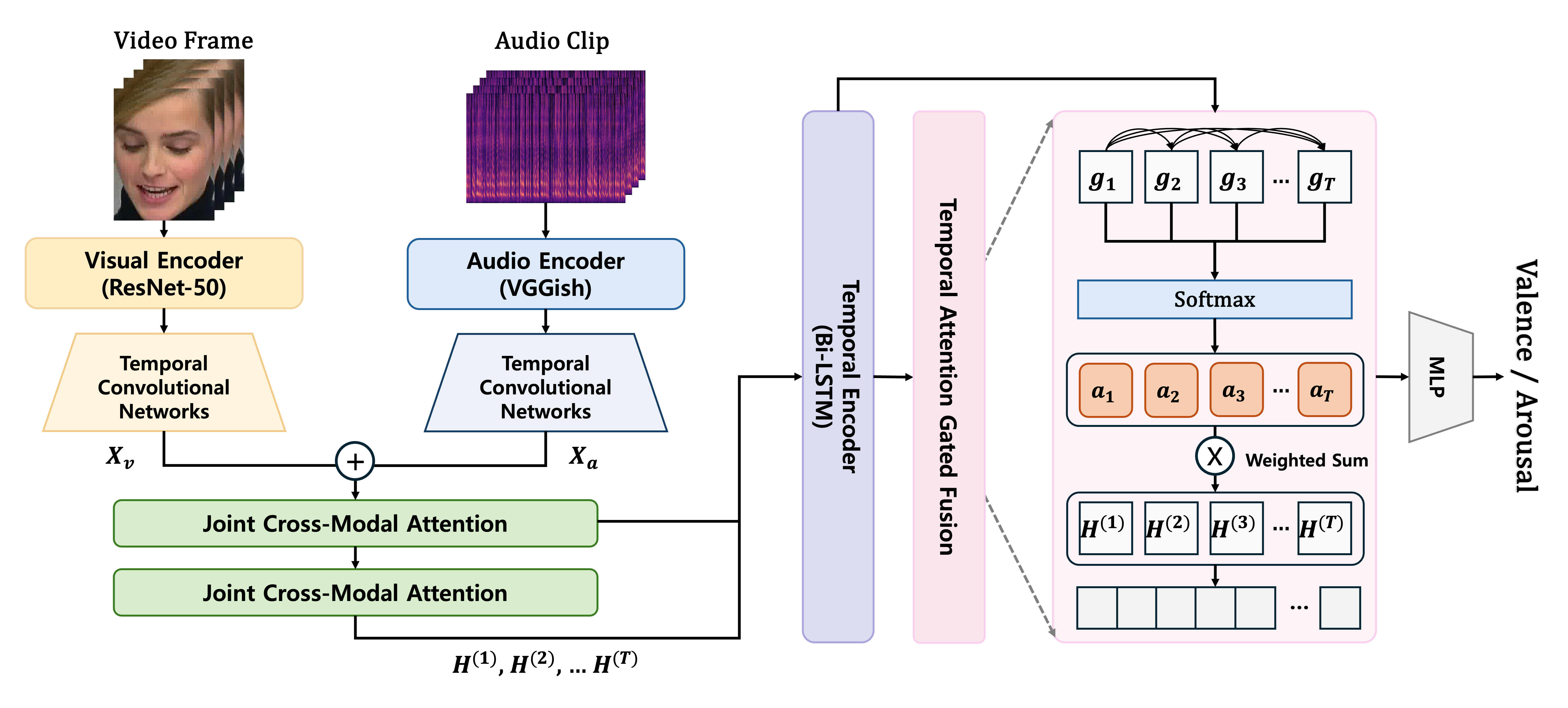}
  \caption{The overall architecture of the proposed TAGF framework. Visual and audio inputs are first encoded via modality-specific backbones and Temporal Convolutional Networks (TCNs), then forwarded through recursive joint cross-modal attention to produce intermediate outputs $\{\mathbf{H}^{(1)}, \dots, \mathbf{H}^{(T)}\}$. These intermediate representations are passed through a Bi-LSTM to capture the temporal dependencies, and time-aware gating weights are computed to aggregate the outputs via weighted summation. The final fused representation is used to predict valence and arousal values through an Multi-Layer Perceptron (MLP).}
  \label{fig:figure2}
  \vspace{-6pt}
\end{figure*}

\section{Methodology}
\label{sec:methodology}

\subsection{Problem Formulation}
The goal of the VA estimation task is to predict continuous emotional values at the frame level from a short video clip. This task can be formulated as a temporal regression problem as follows:
\begin{align}
    \hat{\mathbf{y}}^l_{va} = f_\theta(\mathbf{x}^l_a, \mathbf{x}^l_v), \quad \forall l \in \{1, ..., L\},
\end{align}
where $\hat{\mathbf{y}}^l_{va}$ denotes the predicted emotion vector at frame $l$, which consists of valence $\hat{\mathbf{y}}^l_{val}$ and arousal $\hat{\mathbf{y}}^l_{aro}$ scores. 
The vectors $\mathbf{x}^l_a$ and $\mathbf{x}^l_v$ represent the audio and visual features at frame $l$, respectively, and $f_\theta$ is a regression model parameterized by $\theta$.

The objective of training is to minimize the discrepancy between the predicted and ground-truth emotional values across the entire sequence of length $L$, typically using regression losses such as the Mean Squared Error (MSE) or Concordance Correlation Coefficient (CCC).

To enhance the accuracy and temporal consistency of VA estimation, we propose the TAGF framework. The proposed method introduces a gating mechanism that leverages temporal information across recursive attention steps, thereby enabling the model to effectively capture the gradual and dynamic evolution of emotional states over time.

\subsection{Framework Overview}
Figure~\ref{fig:figure2} illustrates the overall architecture of the proposed TAGF framework for VA estimation. The framework is designed in four stages, which aim to improve the model's ability to capture both temporal emotional dynamics and cross-modal consistency. The details of each stage are described as follows. (\romannumeral1) Given a short video clip, we first extract the frame-level visual and audio features. The visual modality \( \mathbf{X}_v \in \mathbb{R}^{B \times L \times d_v} \) is obtained using a CNN-based encoder, such as ResNet, whereas the audio modality \( \mathbf{X}_a \in \mathbb{R}^{B \times L \times d_a} \) is derived from spectrogram representations processed via a backbone, such as VGGish. Both features are \( \ell_2 \)-normalized as follows:
\begin{align}
    \hat{\mathbf{X}}_v = \text{Normalize}(\mathbf{X}_v), \quad \hat{\mathbf{X}}_a = \text{Normalize}(\mathbf{X}_a)
\end{align}
(\romannumeral2) To model inter-modal correlations, we employ recursive joint cross-attention across \( T \) iterative steps. At each step \( t \), the updated representations are computed as follows:
\begin{align}
    \mathbf{H}_v^{(t)} = \text{CrossAttn}_v(\mathbf{H}_v^{(t-1)}, \mathbf{H}_a^{(t-1)}),
\end{align}
\begin{align}
    \mathbf{H}_a^{(t)} = \text{CrossAttn}_a(\mathbf{H}_a^{(t-1)}, \mathbf{H}_v^{(t-1)}),
\end{align}
where \( \mathbf{H}_v^{(0)} = \hat{\mathbf{X}}_v \) and \( \mathbf{H}_a^{(0)} = \hat{\mathbf{X}}_a \). This process facilitates the progressive refinement of both visual and audio features using joint attention mechanisms. (\romannumeral3) Rather than simply averaging the outputs from different recursion steps, TAGF assigns dynamic weights to each step based on its temporal significance. Specifically, it processes a sequence of recursive outputs using a bidirectional LSTM-based temporal encoder that captures contextual dependencies across steps and produces attention weights for each frame and step. These weights are then used to adaptively fuse the recursive features, allowing the model to emphasize more informative representations while attenuating the noisy or redundant ones. (\romannumeral4) Finally, the gated visual and audio features are concatenated to form a multimodal representation, which is passed through a two-layer MLP regressor to predict the continuous valence and arousal values for each frame. The model is trained using a CCC-based loss function to ensure a robust correlation between the predictions and ground truth:
\begin{align}
    \mathcal{L}_{\text{CCC}} = \sum_{c \in \{v, a\}} \left(1 - \text{CCC}(\mathbf{y}_c, \hat{\mathbf{y}}_c)\right)
\end{align}

In summary, the proposed TAGF framework effectively combines recursive attention modeling with temporal-aware fusion, enabling robust and fine-grained estimation of emotional states in dynamic, real-world settings.

\subsection{Time-aware Gated Fusion}
To effectively address the weak or conflicting correlations between the audio and visual modalities, we propose a TAGF mechanism. Unlike prior studies that uniformly fuse multimodal features over temporal sequences, TAGF adaptively modulates the contribution of each recursive attention output by modeling temporal dependencies. The proposed fusion methodology comprises two principal components: a temporal gated mechanism that learns importance weights dependent on time and a gated feature aggregation module that fuses multi-step attention outputs utilizing these weights.

\subsubsection{Temporal Gated Mechanism}
To effectively incorporate temporal dependencies into the recursive fusion process, we introduce a temporal gated mechanism that learns the relative importance of each attention step based on its temporal context. Let $\{\mathbf{H}^{(1)}, \mathbf{H}^{(2)}, \dots, \mathbf{H}^{(T)}\}$ denote the set of intermediate representations obtained from $T$ recursive cross-attention layers. Each $\mathbf{H}^{(t)} \in \mathbb{R}^{d}$ represents the multimodal fused feature at the $t$-th recursive step.

Rather than treating these outputs independently, we consider them as a temporally ordered sequence and model their contextual interdependencies using a learnable temporal encoder. Specifically, we define a function $f_{\text{temp}}(\cdot)$ that takes a sequence of recursive outputs as input and generates a corresponding sequence of gating vectors:
\begin{align}
\mathbf{G} = f_{\text{temp}}([\mathbf{H}^{(1)}, \mathbf{H}^{(2)}, \dots, \mathbf{H}^{(T)}]) \in \mathbb{R}^{T \times d},
\end{align}
where $\mathbf{G} = \{\mathbf{g}_1, \mathbf{g}_2, \dots, \mathbf{g}_T\}$ and each $\mathbf{g}_t \in \mathbb{R}^{d}$ encodes the temporal context and relative importance of the $t$-th attention output. In our implementation, we adopt a Bidirectional Long Short-Term Memory (Bi-LSTM) network as the temporal encoder $f_{\text{temp}}$, owing to its ability to capture both past and future dependencies in the temporal sequence. The Bi-LSTM processes the input sequence in both forward and backward directions, and the outputs are concatenated to form the final representation at each time step. This design ensures that each gating vector $\mathbf{g}_t$ is aware of its entire temporal context, which is crucial for capturing the dynamic nature of emotion transitions over time.

By treating attention outputs as temporally structured data, this mechanism enables the model to transcend static fusion methodologies and adaptively reweight the contributions of each recursive step depending on the context in which they arise.

\subsubsection{Gated Feature Aggregation}
Once the time-aware gating vectors $\{\mathbf{g}_1, \mathbf{g}_2, \dots, \mathbf{g}_T\}$ are computed, the model proceeds to aggregate the recursive attention outputs $\{\mathbf{H}^{(1)}, \mathbf{H}^{(2)}, \dots, \mathbf{H}^{(T)}\}$ in a weighted manner. Instead of simply averaging or selecting the final step output, we employ a soft gating mechanism to adaptively fuse the recursive features based on their learned relative importance.

Specifically, for each gating vector $\mathbf{g}_t$, we compute a scalar attention score $\alpha_t$ using a dot product with a learnable parameter vector $\mathbf{w} \in \mathbb{R}^{d}$, followed by softmax normalization:
\begin{align}
\alpha_t = \frac{\exp(\mathbf{w}^\top \mathbf{g}_t)}{\sum_{k=1}^{T} \exp(\mathbf{w}^\top \mathbf{g}_k)}
\end{align}
The resulting weights $\{\alpha_1, \dots, \alpha_T\}$ form a probability distribution over the $T$ recursive steps, representing the model's confidence in the relevance of each step's output.

The final fused representation $\mathbf{F} \in \mathbb{R}^{d}$ is computed as the weighted sum of the recursive outputs:
\begin{align}
\mathbf{F} = \sum_{t=1}^{T} \alpha_t \cdot \mathbf{H}^{(t)}
\end{align}
The gated visual and audio features are then concatenated to form the final multimodal representation:
\begin{align}
\mathbf{F}_{\text{multi}} = \text{Concat}(\mathbf{F}_v, \mathbf{F}_a)
\end{align}
The resulting multimodal representation is fed into a two-layer MLP regressor to estimate the frame-wise valence and arousal scores.

The proposed gated aggregation strategy enables the model to place greater emphasis on temporally informative representations while downweighting redundant or noisy outputs. Crucially, because the attention weights $\alpha_t$ are learned in a data-dependent and end-to-end differentiable manner, the model can dynamically adapt to varying emotional dynamics across different video sequences.

Moreover, this mechanism enhances robustness by allowing the model to suppress outputs that may be noisy or temporally misaligned, such as those corresponding to occluded faces or off-screen speech frames, without requiring explicit masking or frame-level supervision. It also avoids the need for heuristic rules or handcrafted temporal smoothing, enabling the model to learn optimal fusion strategies directly from the data.

In summary, the gated feature aggregation module acts as a temporally adaptive attention pooling layer over the recursive cross-modal fusion stages, guided by the learned temporal cues from the Bi-LSTM encoder. Along with the temporal-gated mechanism, this module constitutes a key module within the TAGF framework, responsible for adaptive attention pooling over recursive cross-modal features.

\section{Experiments}
\label{sec:experiments}
\subsection{Datasets}
We trained a VA estimation model using Aff-Wild2~\cite{kollias2019deep}, a benchmark dataset designed for various affective computing tasks such as facial expression recognition, action unit detection, and VA estimation~\cite{zafeiriou2017aff, kollias2019face, kollias2019deep, kollias2019expression, kollias2020analysing, kollias2021distribution, kollias2021affect, kollias2021analysing, kollias2022abaw, kollias2023abaw, kollias2023multi, kollias2023abaw2, kollias20246th, kollias20247th, kollias2025dvd, kolliasadvancements, kollias2024distribution}. Aff-Wild2 is a large-scale dataset constructed for affective computing research, consisting of 598 in-the-wild videos collected from YouTube. Approximately 2,065,871 frames were extracted from 432 unique subjects. Among these, four videos contain two subjects, each annotated separately. The annotations for VA are provided as continuous values within the range of [-1, 1]. The dataset is partitioned into 356 videos for training, 76 for validation, and 162 for testing. The partitioning is subject-independent, ensuring that each subject appears in only one subset, thus allowing the models to generalize to unseen individuals.

\subsection{Implementation Details}
\subsubsection{Preprocessing \& Feature Extraction}
To reflect the temporal characteristics inherent in continuous emotion estimation, this study preprocesses the visual and audio modalities of the Aff-Wild2 dataset with a focus on temporal consistency. The visual inputs consisted of 48 × 48 cropped and aligned face images provided by the challenge organizers. Frames without detected faces or missing annotations are either masked with zeros or excluded from the training. Each video is seperated into sequences of 300 frames with a stride of 200, resulting in an approximately 33\% overlap, which facilitates the learning of gradual emotional transitions. The audio stream is extracted from the original videos, resampled to 16kHz, and converted into log-mel spectrograms following the VGGish~\cite{simonyan2014very} preprocessing pipeline. To ensure precise temporal alignment with the visual stream, the hop length is set to the reciprocal of the video frame rate (1/fps). Both visual and audio inputs are normalized to have a mean of 0.5 and a standard deviation of 0.5 to enhance training stability.

For modality-specific backbones, a ResNet-50~\cite{he2016deep} pretrained on MS-Celeb-1M~\cite{guo2016ms} is employed for the visual stream, whereas a VGG network pretrained on AudioSet is used for the acoustic stream. To capture dynamic temporal patterns, TCNs are applied on top of the frame-level embeddings in both modalities. During training, data augmentation is performed on visual inputs via random flipping and random cropping with a crop size of 40, while only center cropping is applied during validation. The VA are modeled as a separate regression target.

\subsubsection{Experimental Setup}
All experiments were conducted using Python 3.10 and PyTorch 2.5.1 in an environment equipped with an NVIDIA RTX 3080 GPU with 12GB of memory. The model was trained using the Adam optimizer with an initial learning rate of 1e-5, a minimum learning rate of 1e-8, and a weight decay of 0.001. The batch size was set to six, and the training process was capped at a maximum of 100 epochs. Early stopping with a patience of 10 epochs was employed to mitigate the overfitting. A reduceLROnPlateau scheduler was applied to dynamically adjust the learning rate based on the CCC score in the validation set. The backbone networks were initially frozen and progressively fine-tuned in three stages, each involving a warm-up phase followed by a learning rate decay based on performance improvement. The model checkpoint with the highest validation performance was selected for the final evaluation.

\subsection{Evaluation Metrics}
We employ CCC as the primary evaluation metric to quantitatively assess the model’s performance in continuous emotion estimation. The CCC jointly considers the mean, variance, and covariance between the predicted values and ground truth, making it particularly well-suited for tasks involving the continuous dynamics of affective states.

Although the commonly used MSE focuses only on point-wise deviations, it fails to account for temporal alignment or scale differences between the sequences. In contrast, CCC reflects not only the prediction accuracy but also the structural agreement in terms of scale discrepancy, distributional shift, and overall trend consistency, making it a more comprehensive and reliable metric for estimating emotion. The CCC is defined as:
\begin{align}
\text{CCC} = \frac{2 \rho \sigma_{\hat{y}} \sigma_{y}}{\sigma_{\hat{y}}^2 + \sigma_{y}^2 + (\mu_{\hat{y}} - \mu_{y})^2},
\end{align}
where \(\mu_{\hat{y}}\) and \(\sigma_{\hat{y}}^2\) denote the mean and variance of the predictions, \(\mu_{y}\) and \(\sigma_{y}^2\) denote the mean and variance of the ground truth, respectively; and \(\rho\) is the Pearson correlation coefficient between the predicted and true values.

\begin{table}[t]
\centering
\small
\caption{Comparison of CCC performance on the Aff-Wild2 validation set (fold-0), as defined by the official challenge split.}
\label{tab:table1}
\resizebox{0.975\linewidth}{!}{
\begin{tabular}{cccc}
\hline
{\textbf{Model}} & \multicolumn{3}{c}{\textbf{Validation Set}}\\
\cmidrule(lr){2-4}
 & \textbf{Valence CCC} & \textbf{Arousal CCC} & \textbf{CCC Avg} \\
\hline
MM-CV-LC~\cite{zhang2021continuous}   & 0.469 & 0.649 & 0.559 \\
Netease Fuxi Virtual Human~\cite{zhang2023multi}   & 0.464 & 0.640 & 0.552 \\
CtyunAI~\cite{zhou2023leveraging}    & 0.550 & \textbf{0.681} & 0.616 \\
HFUT-MAC~\cite{zhang2023abaw5}   & 0.554 & 0.659 & 0.607 \\
Situ-RUCAIM3~\cite{meng2022valence}    & 0.588 & 0.668 & \textbf{0.628} \\
JCA~\cite{praveen2022joint} & \textbf{0.663} & 0.584 & 0.623 \\
RJCA~\cite{praveen2023recursive} & 0.443 & 0.639 & 0.541 \\
DCA~\cite{praveen2024cross} & 0.451 & 0.647 & 0.549 \\
GRJCA~\cite{mennin2013united} & 0.459 & 0.652 & 0.556 \\
HGRJCA~\cite{mennin2013united} & 0.464 & 0.660 & 0.562 \\
\textbf{Ours}     & 0.427 & 0.676 & 0.552 \\
\hline
\end{tabular}
}
\end{table}

\subsection{VA Estimation Results}
\subsubsection{Performance on the Validation Set}
The proposed TAGF model demonstrates competitive performance in the VA regression task on the Aff-Wild2 dataset compared to recent state-of-the-art audio-visual fusion approaches. The experiments were conducted on the official ABAW validation set (fold-0), and the performance was evaluated using the CCC. As shown in Table~\ref{tab:table1}, the proposed TAGF model achieved CCC of 0.427 and 0.676 for valence and arousal, respectively, resulting in an average CCC of 0.552. Notably, the arousal score of 0.676 ranks second among all the compared methods, closely following that of the top-performing model. Although the average CCC is slightly lower than that of some state-of-the-art approaches, TAGF demonstrates competitive and stable performance despite its relatively simple architecture. Additionally, the model is trained solely on the Aff-Wild2 dataset without relying on multiple pretrained backbones or additional external resources, highlighting its efficiency and robustness.

In particular, the proposed model incorporates a time-aware gating mechanism that effectively captures the dynamic transitions of emotional expressions, thereby enhancing the robustness of the multimodal fusion. Moreover, consistent results observed across cross-validation folds beyond the official split suggest the potential applicability of the model in real-world emotion-recognition scenarios.

\begin{table}[t]
\centering
\small
\caption{Comparison of CCC performance on the Aff-Wild2 test set (fold-0), evaluated through the official challenge server.}
\label{tab:table2}
\resizebox{0.975\linewidth}{!}{
\begin{tabular}{cccc}
\hline
{\textbf{Model}} & \multicolumn{3}{c}{\textbf{Test Set}}\\
\cmidrule(lr){2-4}
 & \textbf{Valence CCC} & \textbf{Arousal CCC} & \textbf{CCC Avg} \\
\hline
MM-CV-LC~\cite{zhang2021continuous}   & 0.463 & 0.492 & 0.478 \\
Netease Fuxi Virtual Human~\cite{zhang2023multi}   & \textbf{0.648} & 0.625 & \textbf{0.637} \\
CtyunAI~\cite{zhou2023leveraging}    & 0.500 & \textbf{0.632} & 0.566 \\
HFUT-MAC~\cite{zhang2023abaw5}   & 0.523 & 0.545 & 0.534 \\
Situ-RUCAIM3~\cite{meng2022valence}    & 0.606 & 0.596 & 0.601 \\
JCA~\cite{praveen2022joint} & 0.374 & 0.363 & 0.369 \\
RJCA~\cite{praveen2023recursive} & 0.537 & 0.576 & 0.557 \\
DCA~\cite{praveen2024cross} & 0.549 & 0.585 & 0.567 \\
GRJCA~\cite{mennin2013united} & 0.556 & 0.605 & 0.581 \\
HGRJCA~\cite{mennin2013united} & 0.561 & 0.620 & 0.591 \\
\textbf{Ours}           & 0.512 & 0.568 & 0.540 \\
\hline
\end{tabular}
}
\end{table}

\subsubsection{Generalization to the Test Set}
To further evaluate the generalization capability of the proposed TAGF model, we submitted the best-performing checkpoint, as determined using the validation set, to the official ABAW challenge server and obtained results on the Aff-Wild2 test set. Table~\ref{tab:table2} compares the performance of the proposed TAGF model with several recent state-of-the-art multimodal approaches evaluated on the test set.

TAGF achieved CCC of 0.512 and 0.568 for valence and arousal, respectively, yielding an average CCC of 0.540. Although the scores are slightly lower than those of top-performing methods such as \citet{zhang2023multi} and \citet{meng2022valence}, the proposed model demonstrates competitive performance with a relatively simple architecture, without relying on external datasets or ensemble techniques.

Moreover, compared with methods such as \citet{praveen2022joint}, TAGF achieves consistent improvements in both valence and arousal estimation. This suggests that the proposed time-aware gating mechanism effectively enhances multimodal fusion and contributes to better generalization under wild conditions.
\section{Conclusion}
\label{sec:conclusion}

In this paper, we propose TAGF, a Time-aware Gated Fusion framework for multimodal emotion recognition, to address the challenge of temporal inconsistency and modality incongruity in real-world valence-arousal estimation tasks. By introducing a Bi-LSTM-based temporal gating mechanism and gated feature aggregation module, TAGF adaptively modulates recursive attention outputs based on their temporal context. This enables the model to better capture the dynamic nature of emotional expressions and the complementary relationship between the visual and audio modalities. Experimental results on the Aff-Wild2 dataset demonstrate that TAGF achieves competitive performance compared to existing recursive attention-based models while exhibiting robustness to noisy or misaligned modalities. These findings highlight the importance of incorporating temporal awareness into multimodal fusion, especially in unconstrained, in-the-wild settings. Future work will explore the integration of modality reliability estimation and uncertainty modeling to further enhance the robustness and interpretability of our proposed framework.
{
    \small
    \bibliographystyle{ieeenat_fullname}
    \bibliography{main}

\begin{thebibliography}{52}
\providecommand{\natexlab}[1]{#1}
\providecommand{\url}[1]{\texttt{#1}}
\expandafter\ifx\csname urlstyle\endcsname\relax
  \providecommand{\doi}[1]{doi: #1}\else
  \providecommand{\doi}{doi: \begingroup \urlstyle{rm}\Url}\fi

\bibitem[Ahire et~al.(2025)Ahire, Shah, Khan, Pakhale, Sookha, Ganaie, and Dhall]{ahire2025maven}
Vrushank Ahire, Kunal Shah, Mudasir Khan, Nikhil Pakhale, Lownish Sookha, Mudasir Ganaie, and Abhinav Dhall.
\newblock Maven: Multi-modal attention for valence-arousal emotion network.
\newblock In \emph{Proceedings of the Computer Vision and Pattern Recognition Conference}, pages 5789--5799, 2025.

\bibitem[Arevalo et~al.(2020)Arevalo, Solorio, Montes-y Gomez, and Gonz{\'a}lez]{arevalo2020gated}
John Arevalo, Thamar Solorio, Manuel Montes-y Gomez, and Fabio~A Gonz{\'a}lez.
\newblock Gated multimodal networks.
\newblock \emph{Neural Computing and Applications}, 32:\penalty0 10209--10228, 2020.

\bibitem[Feldman(1995)]{feldman1995valence}
Lisa~A Feldman.
\newblock Valence focus and arousal focus: Individual differences in the structure of affective experience.
\newblock \emph{Journal of personality and social psychology}, 69\penalty0 (1):\penalty0 153, 1995.

\bibitem[Ghaleb et~al.(2023)Ghaleb, Niehues, and Asteriadis]{ghaleb2023joint}
Esam Ghaleb, Jan Niehues, and Stylianos Asteriadis.
\newblock Joint modelling of audio-visual cues using attention mechanisms for emotion recognition.
\newblock \emph{Multimedia Tools and Applications}, 82\penalty0 (8):\penalty0 11239--11264, 2023.

\bibitem[Guo et~al.(2016)Guo, Zhang, Hu, He, and Gao]{guo2016ms}
Yandong Guo, Lei Zhang, Yuxiao Hu, Xiaodong He, and Jianfeng Gao.
\newblock Ms-celeb-1m: A dataset and benchmark for large-scale face recognition.
\newblock In \emph{Computer Vision--ECCV 2016: 14th European Conference, Amsterdam, The Netherlands, October 11-14, 2016, Proceedings, Part III 14}, pages 87--102. Springer, 2016.

\bibitem[He et~al.(2016)He, Zhang, Ren, and Sun]{he2016deep}
Kaiming He, Xiangyu Zhang, Shaoqing Ren, and Jian Sun.
\newblock Deep residual learning for image recognition.
\newblock In \emph{Proceedings of the IEEE conference on computer vision and pattern recognition}, pages 770--778, 2016.

\bibitem[Hwooi et~al.(2022)Hwooi, Othmani, and Sabri]{hwooi2022deep}
Stephen Khor~Wen Hwooi, Alice Othmani, and Aznul Qalid~Md Sabri.
\newblock Deep learning-based approach for continuous affect prediction from facial expression images in valence-arousal space.
\newblock \emph{IEEE Access}, 10:\penalty0 96053--96065, 2022.

\bibitem[Karas et~al.(2022)Karas, Tellamekala, Mallol-Ragolta, Valstar, and Schuller]{karas2022continuous}
Vincent Karas, Mani~Kumar Tellamekala, Adria Mallol-Ragolta, Michel Valstar, and Bj{\"o}rn~W Schuller.
\newblock Continuous-time audiovisual fusion with recurrence vs. attention for in-the-wild affect recognition.
\newblock \emph{arXiv preprint arXiv:2203.13285}, 2022.

\bibitem[Kollias(2022)]{kollias2022abaw}
Dimitrios Kollias.
\newblock Abaw: Valence-arousal estimation, expression recognition, action unit detection \& multi-task learning challenges.
\newblock In \emph{Proceedings of the IEEE/CVF Conference on Computer Vision and Pattern Recognition}, pages 2328--2336, 2022.

\bibitem[Kollias(2023{\natexlab{a}})]{kollias2023abaw}
Dimitrios Kollias.
\newblock Abaw: Learning from synthetic data \& multi-task learning challenges.
\newblock In \emph{European Conference on Computer Vision}, pages 157--172. Springer, 2023{\natexlab{a}}.

\bibitem[Kollias(2023{\natexlab{b}})]{kollias2023multi}
Dimitrios Kollias.
\newblock Multi-label compound expression recognition: C-expr database \& network.
\newblock In \emph{Proceedings of the IEEE/CVF Conference on Computer Vision and Pattern Recognition}, pages 5589--5598, 2023{\natexlab{b}}.

\bibitem[Kollias and Zafeiriou(2019)]{kollias2019expression}
Dimitrios Kollias and Stefanos Zafeiriou.
\newblock Expression, affect, action unit recognition: Aff-wild2, multi-task learning and arcface.
\newblock \emph{arXiv preprint arXiv:1910.04855}, 2019.

\bibitem[Kollias and Zafeiriou(2021{\natexlab{a}})]{kollias2021affect}
Dimitrios Kollias and Stefanos Zafeiriou.
\newblock Affect analysis in-the-wild: Valence-arousal, expressions, action units and a unified framework.
\newblock \emph{arXiv preprint arXiv:2103.15792}, 2021{\natexlab{a}}.

\bibitem[Kollias and Zafeiriou(2021{\natexlab{b}})]{kollias2021analysing}
Dimitrios Kollias and Stefanos Zafeiriou.
\newblock Analysing affective behavior in the second abaw2 competition.
\newblock In \emph{Proceedings of the IEEE/CVF International Conference on Computer Vision}, pages 3652--3660, 2021{\natexlab{b}}.

\bibitem[Kollias et~al.()Kollias, Schulc, Hajiyev, and Zafeiriou]{kollias2020analysing}
D Kollias, A Schulc, E Hajiyev, and S Zafeiriou.
\newblock Analysing affective behavior in the first abaw 2020 competition.
\newblock In \emph{2020 15th IEEE International Conference on Automatic Face and Gesture Recognition (FG 2020)(FG)}, pages 794--800.

\bibitem[Kollias et~al.(2019{\natexlab{a}})Kollias, Sharmanska, and Zafeiriou]{kollias2019face}
Dimitrios Kollias, Viktoriia Sharmanska, and Stefanos Zafeiriou.
\newblock Face behavior a la carte: Expressions, affect and action units in a single network.
\newblock \emph{arXiv preprint arXiv:1910.11111}, 2019{\natexlab{a}}.

\bibitem[Kollias et~al.(2019{\natexlab{b}})Kollias, Tzirakis, Nicolaou, Papaioannou, Zhao, Schuller, Kotsia, and Zafeiriou]{kollias2019deep}
Dimitrios Kollias, Panagiotis Tzirakis, Mihalis~A Nicolaou, Athanasios Papaioannou, Guoying Zhao, Bj{\"o}rn Schuller, Irene Kotsia, and Stefanos Zafeiriou.
\newblock Deep affect prediction in-the-wild: Aff-wild database and challenge, deep architectures, and beyond.
\newblock \emph{International Journal of Computer Vision}, pages 1--23, 2019{\natexlab{b}}.

\bibitem[Kollias et~al.(2021)Kollias, Sharmanska, and Zafeiriou]{kollias2021distribution}
Dimitrios Kollias, Viktoriia Sharmanska, and Stefanos Zafeiriou.
\newblock Distribution matching for heterogeneous multi-task learning: a large-scale face study.
\newblock \emph{arXiv preprint arXiv:2105.03790}, 2021.

\bibitem[Kollias et~al.(2023)Kollias, Tzirakis, Baird, Cowen, and Zafeiriou]{kollias2023abaw2}
Dimitrios Kollias, Panagiotis Tzirakis, Alice Baird, Alan Cowen, and Stefanos Zafeiriou.
\newblock Abaw: Valence-arousal estimation, expression recognition, action unit detection \& emotional reaction intensity estimation challenges.
\newblock In \emph{Proceedings of the IEEE/CVF Conference on Computer Vision and Pattern Recognition}, pages 5888--5897, 2023.

\bibitem[Kollias et~al.(2024{\natexlab{a}})Kollias, Sharmanska, and Zafeiriou]{kollias2024distribution}
Dimitrios Kollias, Viktoriia Sharmanska, and Stefanos Zafeiriou.
\newblock Distribution matching for multi-task learning of classification tasks: a large-scale study on faces \& beyond.
\newblock In \emph{Proceedings of the AAAI Conference on Artificial Intelligence}, pages 2813--2821, 2024{\natexlab{a}}.

\bibitem[Kollias et~al.(2024{\natexlab{b}})Kollias, Tzirakis, Cowen, Zafeiriou, Kotsia, Baird, Gagne, Shao, and Hu]{kollias20246th}
Dimitrios Kollias, Panagiotis Tzirakis, Alan Cowen, Stefanos Zafeiriou, Irene Kotsia, Alice Baird, Chris Gagne, Chunchang Shao, and Guanyu Hu.
\newblock The 6th affective behavior analysis in-the-wild (abaw) competition.
\newblock In \emph{Proceedings of the IEEE/CVF Conference on Computer Vision and Pattern Recognition}, pages 4587--4598, 2024{\natexlab{b}}.

\bibitem[Kollias et~al.(2024{\natexlab{c}})Kollias, Zafeiriou, Kotsia, Dhall, Ghosh, Shao, and Hu]{kollias20247th}
Dimitrios Kollias, Stefanos Zafeiriou, Irene Kotsia, Abhinav Dhall, Shreya Ghosh, Chunchang Shao, and Guanyu Hu.
\newblock 7th abaw competition: Multi-task learning and compound expression recognition.
\newblock \emph{arXiv preprint arXiv:2407.03835}, 2024{\natexlab{c}}.

\bibitem[Kollias et~al.(2025{\natexlab{a}})Kollias, Senadeera, Zheng, Yadav, Slabaugh, Awais, and Yang]{kollias2025dvd}
Dimitrios Kollias, Damith~C Senadeera, Jianian Zheng, Kaushal~KK Yadav, Greg Slabaugh, Muhammad Awais, and Xiaoyun Yang.
\newblock Dvd: A comprehensive dataset for advancing violence detection in real-world scenarios.
\newblock \emph{arXiv preprint arXiv:2506.05372}, 2025{\natexlab{a}}.

\bibitem[Kollias et~al.(2025{\natexlab{b}})Kollias, Tzirakis, Cowen, Kotsia, Cogitat, Granger, Pedersoli, Bacon, Baird, Shao, et~al.]{kolliasadvancements}
Dimitrios Kollias, Panagiotis Tzirakis, Alan Cowen, Irene Kotsia, UK Cogitat, Eric Granger, Marco Pedersoli, Simon Bacon, Alice Baird, Chunchang Shao, et~al.
\newblock Advancements in affective and behavior analysis: The 8th abaw workshop and competition.
\newblock 2025{\natexlab{b}}.

\bibitem[Kumar and Vepa(2020)]{kumar2020gated}
Ayush Kumar and Jithendra Vepa.
\newblock Gated mechanism for attention based multi modal sentiment analysis.
\newblock In \emph{ICASSP 2020-2020 IEEE International Conference on Acoustics, Speech and Signal Processing (ICASSP)}, pages 4477--4481. IEEE, 2020.

\bibitem[LaGrandeur(2015)]{lagrandeur2015emotion}
Kevin LaGrandeur.
\newblock Emotion, artificial intelligence, and ethics.
\newblock \emph{Beyond artificial intelligence: The disappearing human-machine divide}, pages 97--109, 2015.

\bibitem[Liu et~al.(2023)Liu, Gao, Li, Fu, and Ding]{liu2023multi}
Shuai Liu, Peng Gao, Yating Li, Weina Fu, and Weiping Ding.
\newblock Multi-modal fusion network with complementarity and importance for emotion recognition.
\newblock \emph{Information Sciences}, 619:\penalty0 679--694, 2023.

\bibitem[Meng et~al.(2022)Meng, Liu, Liu, Huang, Jiang, Zhang, Liu, and Jin]{meng2022valence}
Liyu Meng, Yuchen Liu, Xiaolong Liu, Zhaopei Huang, Wenqiang Jiang, Tenggan Zhang, Chuanhe Liu, and Qin Jin.
\newblock Valence and arousal estimation based on multimodal temporal-aware features for videos in the wild.
\newblock In \emph{Proceedings of the IEEE/CVF Conference on Computer Vision and Pattern Recognition}, pages 2345--2352, 2022.

\bibitem[Mennin et~al.(2013)Mennin, Ellard, Fresco, and Gross]{mennin2013united}
Douglas~S Mennin, Kristen~K Ellard, David~M Fresco, and James~J Gross.
\newblock United we stand: Emphasizing commonalities across cognitive-behavioral therapies.
\newblock \emph{Behavior therapy}, 44\penalty0 (2):\penalty0 234--248, 2013.

\bibitem[Ming et~al.(2022)Ming, Qian, and Guangyuan]{ming2022cnn}
Ye Ming, Hu Qian, and Liu Guangyuan.
\newblock Cnn-lstm facial expression recognition method fused with two-layer attention mechanism.
\newblock \emph{Computational Intelligence and Neuroscience}, 2022\penalty0 (1):\penalty0 7450637, 2022.

\bibitem[Mocanu et~al.(2023)Mocanu, Tapu, and Zaharia]{mocanu2023multimodal}
Bogdan Mocanu, Ruxandra Tapu, and Titus Zaharia.
\newblock Multimodal emotion recognition using cross modal audio-video fusion with attention and deep metric learning.
\newblock \emph{Image and Vision Computing}, 133:\penalty0 104676, 2023.

\bibitem[Praveen and Alam(2024{\natexlab{a}})]{praveen2024cross}
R~Gnana Praveen and Jahangir Alam.
\newblock Cross-attention is not always needed: Dynamic cross-attention for audio-visual dimensional emotion recognition.
\newblock In \emph{2024 IEEE International Conference on Multimedia and Expo (ICME)}, pages 1--6. IEEE, 2024{\natexlab{a}}.

\bibitem[Praveen and Alam(2024{\natexlab{b}})]{praveen2024incongruity}
R~Gnana Praveen and Jahangir Alam.
\newblock Incongruity-aware cross-modal attention for audio-visual fusion in dimensional emotion recognition.
\newblock \emph{IEEE Journal of Selected Topics in Signal Processing}, 2024{\natexlab{b}}.

\bibitem[Praveen and Alam(2024{\natexlab{c}})]{praveen2024recursive}
R~Gnana Praveen and Jahangir Alam.
\newblock Recursive joint cross-modal attention for multimodal fusion in dimensional emotion recognition.
\newblock In \emph{Proceedings of the IEEE/CVF Conference on Computer Vision and Pattern Recognition}, pages 4803--4813, 2024{\natexlab{c}}.

\bibitem[Praveen et~al.(2022)Praveen, de~Melo, Ullah, Aslam, Zeeshan, Denorme, Pedersoli, Koerich, Bacon, Cardinal, et~al.]{praveen2022joint}
R~Gnana Praveen, Wheidima~Carneiro de Melo, Nasib Ullah, Haseeb Aslam, Osama Zeeshan, Th{\'e}o Denorme, Marco Pedersoli, Alessandro~L Koerich, Simon Bacon, Patrick Cardinal, et~al.
\newblock A joint cross-attention model for audio-visual fusion in dimensional emotion recognition.
\newblock In \emph{Proceedings of the IEEE/CVF conference on computer vision and pattern recognition}, pages 2486--2495, 2022.

\bibitem[Praveen et~al.(2023{\natexlab{a}})Praveen, Cardinal, and Granger]{praveen2023audio}
R~Gnana Praveen, Patrick Cardinal, and Eric Granger.
\newblock Audio--visual fusion for emotion recognition in the valence--arousal space using joint cross-attention.
\newblock \emph{IEEE Transactions on Biometrics, Behavior, and Identity Science}, 5\penalty0 (3):\penalty0 360--373, 2023{\natexlab{a}}.

\bibitem[Praveen et~al.(2023{\natexlab{b}})Praveen, Granger, and Cardinal]{praveen2023recursive}
R~Gnana Praveen, Eric Granger, and Patrick Cardinal.
\newblock Recursive joint attention for audio-visual fusion in regression based emotion recognition.
\newblock In \emph{ICASSP 2023-2023 IEEE International Conference on Acoustics, Speech and Signal Processing (ICASSP)}, pages 1--5. IEEE, 2023{\natexlab{b}}.

\bibitem[Russell(1980)]{russell1980circumplex}
James~A Russell.
\newblock A circumplex model of affect.
\newblock \emph{Journal of personality and social psychology}, 39\penalty0 (6):\penalty0 1161, 1980.

\bibitem[S{\'a}nchez-Lozano et~al.(2013)S{\'a}nchez-Lozano, Lopez-Otero, Docio-Fernandez, Argones-R{\'u}a, and Alba-Castro]{sanchez2013audiovisual}
Enrique S{\'a}nchez-Lozano, Paula Lopez-Otero, Laura Docio-Fernandez, Enrique Argones-R{\'u}a, and Jos{\'e}~Luis Alba-Castro.
\newblock Audiovisual three-level fusion for continuous estimation of russell's emotion circumplex.
\newblock In \emph{Proceedings of the 3rd ACM international workshop on Audio/visual emotion challenge}, pages 31--40, 2013.

\bibitem[Sharafi et~al.(2022)Sharafi, Yazdchi, Rasti, and Nasimi]{sharafi2022novel}
Masoumeh Sharafi, Mohammadreza Yazdchi, Reza Rasti, and Fahimeh Nasimi.
\newblock A novel spatio-temporal convolutional neural framework for multimodal emotion recognition.
\newblock \emph{Biomedical Signal Processing and Control}, 78:\penalty0 103970, 2022.

\bibitem[Simonyan and Zisserman(2014)]{simonyan2014very}
Karen Simonyan and Andrew Zisserman.
\newblock Very deep convolutional networks for large-scale image recognition.
\newblock \emph{arXiv preprint arXiv:1409.1556}, 2014.

\bibitem[Tripathi et~al.(2017)Tripathi, Acharya, Sharma, Mittal, and Bhattacharya]{tripathi2017using}
Samarth Tripathi, Shrinivas Acharya, Ranti Sharma, Sudhanshi Mittal, and Samit Bhattacharya.
\newblock Using deep and convolutional neural networks for accurate emotion classification on deap data.
\newblock In \emph{Proceedings of the AAAI Conference on Artificial Intelligence}, pages 4746--4752, 2017.

\bibitem[Xie et~al.(2023)Xie, Wang, Wang, Yang, Gu, Tang, and Varatnitski]{xie2023multimodal}
Jinbao Xie, Jiyu Wang, Qingyan Wang, Dali Yang, Jinming Gu, Yongqiang Tang, and Yury~I Varatnitski.
\newblock A multimodal fusion emotion recognition method based on multitask learning and attention mechanism.
\newblock \emph{Neurocomputing}, 556:\penalty0 126649, 2023.

\bibitem[Xie et~al.(2019)Xie, Liang, Liang, Huang, Zou, and Schuller]{xie2019speech}
Yue Xie, Ruiyu Liang, Zhenlin Liang, Chengwei Huang, Cairong Zou, and Bj{\"o}rn Schuller.
\newblock Speech emotion classification using attention-based lstm.
\newblock \emph{IEEE/ACM Transactions on Audio, Speech, and Language Processing}, 27\penalty0 (11):\penalty0 1675--1685, 2019.

\bibitem[Zafeiriou et~al.(2017)Zafeiriou, Kollias, Nicolaou, Papaioannou, Zhao, and Kotsia]{zafeiriou2017aff}
Stefanos Zafeiriou, Dimitrios Kollias, Mihalis~A Nicolaou, Athanasios Papaioannou, Guoying Zhao, and Irene Kotsia.
\newblock Aff-wild: Valence and arousal ‘in-the-wild’challenge.
\newblock In \emph{Computer Vision and Pattern Recognition Workshops (CVPRW), 2017 IEEE Conference on}, pages 1980--1987. IEEE, 2017.

\bibitem[Zhang et~al.(2021)Zhang, Ding, Wei, and Guan]{zhang2021continuous}
Su Zhang, Yi Ding, Ziquan Wei, and Cuntai Guan.
\newblock Continuous emotion recognition with audio-visual leader-follower attentive fusion.
\newblock In \emph{Proceedings of the IEEE/CVF international conference on computer vision}, pages 3567--3574, 2021.

\bibitem[Zhang et~al.(2024)Zhang, Yang, Chen, Zhang, Leng, and Zhao]{zhang2024deep}
Shiqing Zhang, Yijiao Yang, Chen Chen, Xingnan Zhang, Qingming Leng, and Xiaoming Zhao.
\newblock Deep learning-based multimodal emotion recognition from audio, visual, and text modalities: A systematic review of recent advancements and future prospects.
\newblock \emph{Expert Systems with Applications}, 237:\penalty0 121692, 2024.

\bibitem[Zhang et~al.(2025)Zhang, Liu, Jiao, Zhang, Chen, and Li]{zhang2025multimodal}
Shunxiang Zhang, Jiajia Liu, Yixuan Jiao, Yulei Zhang, Lei Chen, and Kuanching Li.
\newblock A multimodal semantic fusion network with cross-modal alignment for multimodal sentiment analysis.
\newblock \emph{ACM Transactions on Multimedia Computing, Communications and Applications}, 2025.

\bibitem[Zhang et~al.(2023{\natexlab{a}})Zhang, Ma, Qiu, and Ding]{zhang2023multi}
Wei Zhang, Bowen Ma, Feng Qiu, and Yu Ding.
\newblock Multi-modal facial affective analysis based on masked autoencoder.
\newblock In \emph{Proceedings of the IEEE/CVF Conference on Computer Vision and Pattern Recognition}, pages 5793--5802, 2023{\natexlab{a}}.

\bibitem[Zhang et~al.(2020)Zhang, Huang, Zeng, and Shan]{zhang2020m}
Yuan-Hang Zhang, Rulin Huang, Jiabei Zeng, and Shiguang Shan.
\newblock M 3 f: Multi-modal continuous valence-arousal estimation in the wild.
\newblock In \emph{2020 15th IEEE International Conference on Automatic Face and Gesture Recognition (FG 2020)}, pages 632--636. IEEE, 2020.

\bibitem[Zhang et~al.(2023{\natexlab{b}})Zhang, An, Cui, Xu, Dong, Jiang, Shi, Liu, Sun, and Wang]{zhang2023abaw5}
Ziyang Zhang, Liuwei An, Zishun Cui, Ao Xu, Tengteng Dong, Yueqi Jiang, Jingyi Shi, Xin Liu, Xiao Sun, and Meng Wang.
\newblock Abaw5 challenge: A facial affect recognition approach utilizing transformer encoder and audiovisual fusion.
\newblock In \emph{Proceedings of the IEEE/CVF Conference on Computer Vision and Pattern Recognition}, pages 5725--5734, 2023{\natexlab{b}}.

\bibitem[Zhou et~al.(2023)Zhou, Lu, Xiong, and Wang]{zhou2023leveraging}
Weiwei Zhou, Jiada Lu, Zhaolong Xiong, and Weifeng Wang.
\newblock Leveraging tcn and transformer for effective visual-audio fusion in continuous emotion recognition.
\newblock In \emph{Proceedings of the IEEE/CVF Conference on Computer Vision and Pattern Recognition}, pages 5756--5763, 2023.

\end{thebibliography}
}


\end{document}